\documentclass[prl,aps,10pt,showpacs,preprintnumbers,amsmath,amssymb,tightenlines,twocolumn]{revtex4-2}

\usepackage{graphicx}
\usepackage{dcolumn}
\usepackage{bm}
\usepackage{epsfig}
\usepackage{amsmath}

\newcommand{\ssection}[1]{{\noi  \it #1:}}

\newcommand{\ket}[1]{|\,{#1}\,\rangle}

\newcommand{\sub}[2]{{#1}_{\mbox{\!\! \scriptsize #2}}}

\def\noi{\noindent}
\def\beq{\begin{equation}}
	\def\eeq{\end{equation}}

\newcommand{\fref}[1]{Fig.~\ref{#1}}

\newcommand{\eref}[1]{Eq.~(\ref{#1})}

\newcommand{\cref}[1]{chapter~\ref{#1}}

\newcommand{\Cref}[1]{Chapter~\ref{#1}}

\newcommand{\Tref}[1]{Table~\ref{#1}}

\newcommand{\bref}[1]{(\ref{#1})}

\usepackage{ulem}  
\usepackage{color}

\begin{document}
	\title{Programmable glassy dynamics using tunable disorder in tweezer arrays}
	\author{K.~Mukherjee}
	\affiliation{Homer L. Dodge Department of Physics and Astronomy, The University of Oklahoma, Norman, Oklahoma 73019, USA}
	\affiliation{Center for Quantum Research and Technology, The University of Oklahoma, Norman, Oklahoma 73019, USA}
	\author{G.~W.~Biedermann}
	\affiliation{Homer L. Dodge Department of Physics and Astronomy, The University of Oklahoma, Norman, Oklahoma 73019, USA}
	\affiliation{Center for Quantum Research and Technology, The University of Oklahoma, Norman, Oklahoma 73019, USA}
	\author{R.~J.~Lewis-Swan}
	\affiliation{Homer L. Dodge Department of Physics and Astronomy, The University of Oklahoma, Norman, Oklahoma 73019, USA}
	\affiliation{Center for Quantum Research and Technology, The University of Oklahoma, Norman, Oklahoma 73019, USA}
	\email{lewisswan@ou.edu}
	\begin{abstract}
        We propose a unifying framework for non-equilibrium relaxation dynamics in ensembles of positionally disordered interacting quantum spins based on the statistical properties, such as mean and  variance, of the underlying disorder distribution.  
        Our framework is validated through extensive exact numerical calculations and we use it to disentangle and understand the importance of dimensionality and interaction range for the observation of glassy (i.e., sub-exponential) decay dynamics. Leveraging the deterministic control of qubit positioning enabled by modern tweezer array architectures, we also introduce a method (``J-mapping'') that can be used to emulate the relaxation dynamics of a disordered system with arbitrary dimensionality and interaction range in bespoke one-dimensional arrays. Our approach paves the way towards tunable relaxation dynamics that can be explored in quantum simulators based on arrays of neutral atoms and molecules.

	\end{abstract}

	\maketitle
	
	\ssection{Introduction}
The dynamics of disordered quantum many-body systems is of fundamental and practical interest across various fields, including solid state \cite{vojta2019disorder}, atomic and molecular physics \cite{schreiber2015observation,signoles2021glassy,franz2022observation,hazzard2014many}, biophysics \cite{scholak2011efficient,plenio2008dephasing} and quantum chemistry \cite{wellnitz2022disorder,rebentrost2009environment}. 
In particular, understanding the influence of disorder on relaxation dynamics, such as the slowdown or even full arrest of thermalization of isolated many-body systems, is a key challenge.

Recent advances in single- and many-body control of atomic and molecular quantum simulators have enabled detailed studies of a range of phenomena in disordered many-body systems, including spin glasses \cite{binder1986spin,phillips1996stretched}, many-body localization \cite{kjall2014many}, thermalization \cite{eisert2015quantum} and time-crystalline phases \cite{zhang2017observation,choi2017observation}.
The remarkable maturation of optical tweezers for trapping individual atoms \cite{nogrette2014single,barredo2016atom,endres2016atom,barredo2018synthetic,manetsch2024tweezer} and molecules  \cite{ospelkaus2006ultracold,ni2018dipolar,yan2013observation,hazzard2014many} 
has led to the demonstration of (dynamic) programmable arrays of atomic and molecular qubits for quantum simulation \cite{weimer2010rydberg,bernien2017probing,levine2018high,browaeys2020many,kaufman2012cooling,madjarov2020high,scholl2021quantum,mukherjee2020two,morgado2021quantum,kim2018detailed,mukherjee2024excitons,Zhang2024} and computation \cite{cohen2021quantum,wurtz2023aquila, Mitra2023}. 
This ability to arbitrarily position individual atoms can provide a powerful tool to investigate many-body systems featuring positional or interaction (bond) disorder. 

Previous experimental investigations of positionally disordered systems have focused on the slow relaxation of collective order, termed ``glassy'' relaxation, and have included frozen Rydberg gases \cite{signoles2021glassy,franz2022observation}, polar molecules in optical lattices \cite{doi:10.1126/science.adf8999,holland2023demand,ruttley2023formation,vilas2024optical}, and nitrogen-vacancy (NV) centers in diamond \cite{kucsko2018critical,davis2023probing,zu2021emergent,hughes2024strongly}. 
Each platform features naturally arising 
positional disorder, though the degree of control is limited to coarse properties, such as the array filling fraction or particle density. Tweezer arrays of neutral atoms \cite{barredo2016atom,endres2016atom,barredo2018synthetic,manetsch2024tweezer} or molecules \cite{ospelkaus2006ultracold,ni2018dipolar,yan2013observation,hazzard2014many}, on the other hand, can provide deterministic control of the positional disorder and flexibility in statistics of the implemented disorder. These technical capabilities can bootstrap the understanding of relaxation dynamics beyond existing cluster-based models, which explain the slow relaxation of collective order as arising from the dephasing of small clusters of strongly coupled spins. These insights can be leveraged to develop tools for 
programmable control of relaxation dynamics and the propagation of quantum information in many-body systems.

With this perspective, here we develop a unifying framework of glassy dynamics in spin models based on the statistics of positional disorder. Specifically, we use extensive numerical simulations of a range of spin models to establish a framework to characterize relaxation dynamics, and in which disorder is disentangled from dimensionality and interaction range. Thus, a key objective is to identify the fundamental factors governing slow relaxation, particularly: (1) the statistics of the disorder, (2) the nature of interactions, and (3) the dimensionality of the system. 
We leverage our framework to extend our investigation beyond 
randomly filled dilute arrays examined in previous works. As an example, we discuss how relaxation in dilute $2$D and $3$D systems can be emulated in bespoke 1D arrays. 
We also outline an alternative approach to control relaxation dynamics using anisotropic dipolar interactions. 

Our work is timely and relevant given the capabilities of state-of-the-art tweezer array platforms. Moreover, our study paves the way for programmable glassy dynamics in disordered systems, with the potential for diverse applications including quantum batteries \cite{ghosh2020enhancement} and highly efficient solid state devices \cite{haripriya2023interface}, which require slow and tunable relaxation to incur minimal loss of energy. 

\begin{figure}[tb]
	\centering
	\epsfig{file=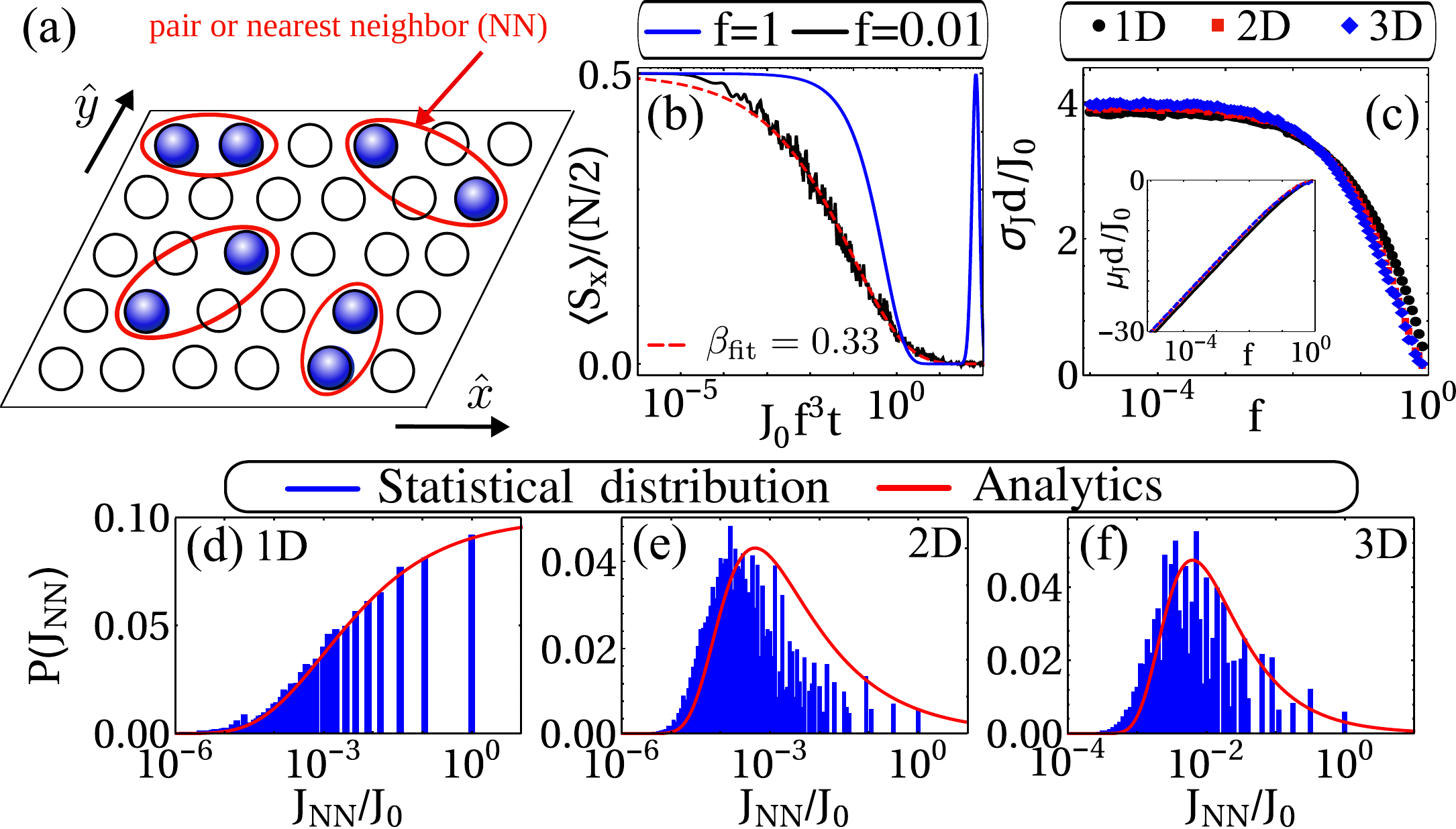,width= 1.05\linewidth}
	\caption{(a) Schematic diagram of a 2D disordered system realised with a dilute array of $N$ qubits (blue-balls) distributed stochastically in a lattice of $L^2$ sites. (b) Relaxation dynamics in a 1D dilute array with Ising interactions (i.e.~$\Delta\gg1$) and $\alpha=3$ at filling fractions $f=0.001$ (black line) and $f=1$ (blue line). The red dashed line indicates a fitted stretched exponential [see Eq.~\bref{Sx_equation}] (c) Rescaled standard deviation $\sigma_J d /J_0$ of the distribution of pair interactions for power-law exponent $\alpha=3$, shown in (d-f), with varying filling fraction in 1D (black), 2D (red), and 3D (blue) dilute arrays. Rescaled mean $\mu_J d/J_0$ is shown in the inset. Distribution of nearest-neighbor interactions in (d) 1D with $f=0.05$, (e) 2D and (f) 3D with $f=0.001$. The red line indicates an approximate analytical expression describing the distribution $P(J)$, see SM for more details. 
		\label{Fig1}}
\end{figure}
%
\ssection{Dynamics of disordered XXZ model}\label{sec:disordered_XXZ} 
As a general basis for our investigation, we consider the dynamics of $N$ qubits described by the Heisenberg XXZ Hamiltonian ($\hbar=1)$, 
\begin{equation}
	\sub{H}{XXZ} = \sum_{i>j} J_{ij}(\hat{S}^{x}_{i} \hat{S}^{x}_{j} + \hat{S}^{y}_{i} \hat{S}^{y}_{j} + \Delta \hat{S}^{z}_{i} \hat{S}^{z}_{j} ) ,
	\label{Hamil_XXZ_1}
\end{equation}
where $\Delta$ is the anisotropy parameter and $\hat{S}^a_{k}$ for $a\in\{x,y,z\}$ are Pauli matrices acting on the $k$th qubit. The qubits interact according to $J_{ij}=J_0/|\mathbf{r}_{ij}|^\alpha$ where $|\mathbf{r}_{ij}|$ is the distance between qubits $i$ and $j$. Here $J_0=C_\alpha/a_0^\alpha$ with $a_0$ a constant with units of distance (e.g., lattice constant for a regular array), and $C_\alpha$ defines the interaction strength for a power-law exponent $\alpha$. 

A starting point for our investigation is the touchstone problem of a dilute $d$-dimensional array with $L^d$ equally spaced sites (with $L$ the side length of the array). This system can be realized by stochastically positioning $N$ qubits, represented by individual tweezers with neutral atoms or molecules, on a lattice of possible locations, with a filling fraction of $f = N/L^d$ [see \fref{Fig1}~(a)].
The filling fraction tunes the disorder strength: $f = 1$ indicates a regular array while $f\to0$ is strongly disordered \cite{aramthottil2024phenomenology}. 
The XXZ Hamiltonian can be realized by, e.g., encoding the qubit states $\ket{\uparrow}$ and $\ket{\downarrow}$ in either Rydberg \cite{nogrette2014single,barredo2016atom,endres2016atom,barredo2018synthetic} or molecular \cite{ospelkaus2006ultracold,ni2018dipolar,yan2013observation,hazzard2014many} states with different parity.


We study the quench dynamics for an initial collective state $|+X\rangle = \big[1/\sqrt{2}(\ket{\uparrow}+\ket{\downarrow})\big]^{\otimes N}$. The time evolution of the collective magnetization $\langle \hat{S}_x (t)\rangle=\sum_j \langle \hat{S}^x_j \rangle$ is obtained by exact numerical integration of the Schr{\"o}dinger equation and averaging over disorder realizations. While our approach is limited to a small number of qubits, it is sufficient to obtain converged results in the limit of $f\rightarrow0$ as the self-averaging of a large system is effectively replicated by taking many disorder realizations. 

The relaxation of the collective magnetization, as shown in \fref{Fig1}~(b), is well approximated by a stretched exponential
\begin{equation}
	\langle \hat{S}_x \rangle = 0.5 \exp[-(\gamma t)^\beta],
    \label{Sx_equation}
\end{equation}
where $\gamma$ is the relaxation rate and $\beta$ is the stretch exponent. In \fref{Fig1}~(b), we show an example time trace of the collective magnetization in a 1D dilute array with $\Delta \gg 1$ and $\alpha=3$. For an ordered array ($f=1$, blue-line), fast relaxation with $\beta=2$ is observed, whereas for a disordered array ($f=0.01$, black-line), we observe slow relaxation with $\beta=0.33$. In general, relaxation characterized by sub-exponential decay of the collective magnetization, i.e., $\beta < 1$, is identified as \textit{glassy}. 

In the strongly disordered limit, $f \ll 1$, the dynamics of $\langle \hat{S}_x (t)\rangle$ can be understood by treating small clusters of strongly interacting spins \cite{schultzen2022glassy,schultzen2022semiclassical,signoles2021glassy,mukherjee2024influence}.
In the simplest model of two spins -- dubbed a \textit{pair model} -- the magnetization dynamics is described by $\langle \hat{S}_x \rangle_{\rm pair} = \cos(\delta \sub{J}{NN}t)$, where $J_{\rm NN}$ denotes the nearest-neighbor (pair) interaction strength and $\delta$ is a model dependent constant \cite{franz2022observation}. 
In the Ising limit ($\Delta\gg1$), summing over all pairs in the array (i.e., averaging over $J_{\rm NN}$) yields a prediction $\sub{\beta}{pair}=d/\alpha$ for a $d$-dimensional array and power-law exponent $\alpha$. Thus, glassy dynamics $(\beta < 1)$ is expected for $d < \alpha$.

While the pair model provides an exact prediction for the Ising limit, its quantitative applicability for other models requires careful analysis \cite{mukherjee2024influence}. Thus, a broader framework connecting key characteristics of disorder statistics with glassy dynamics via independently tunable parameters, such as dimensionality and interaction range, can provide a foundational understanding of glassy dynamics beyond dilute arrays. 


%
%
 \ssection{Role of disorder statistics}\label{sec:stretched_exp} 
\begin{figure}[tb]
	\centering
	\epsfig{file=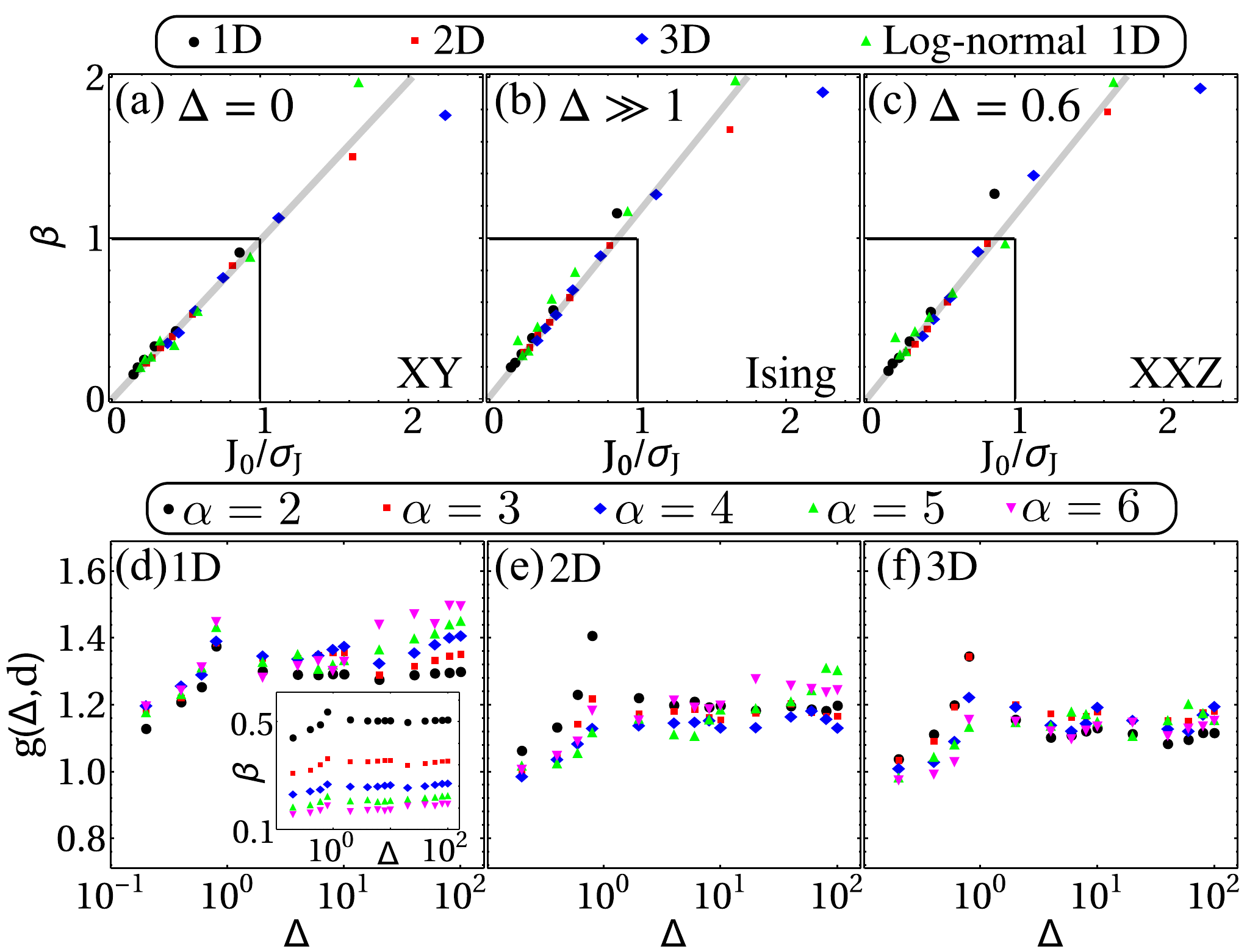,width= \linewidth}
	\caption{Variation in stretch exponent $\beta$ with inverse standard deviation $J_0/\sigma_J$ for (a) XY ($\Delta=0$), (b) Ising ($\Delta\gg 1$) and (c) XXZ ($\Delta=0.6$) models. Data is obtained by varying the power-law exponent between $\alpha = 2...6$. (d-f) Behaviour of prefactor in $g(\Delta,d)=\beta \sigma_J/J_0$ with model anisotropy $\Delta$ in (d) 1D, (e) 2D and (f) 3D disordered arrays for different power-law exponents $\alpha$. Inset shows the comparative variation of $\beta$ with $\Delta$ in 1D before rescaling by $\sigma_J$. The filling fraction is fixed at $f=0.01$ for all cases.
		\label{Fig2}}
\end{figure}
To elucidate the ingredients for slow relaxation, we analyze the distribution of pair or nearest-neighbor (NN) interactions in a dilute array. 
\fref{Fig1}~(d-f) show $P(J)$ for 1D, 2D and 3D arrays in the limit of $f\rightarrow0$ and fixed $\alpha = 3$, for which glassy relaxation has previously been investigated \cite{signoles2021glassy,schultzen2022glassy,tan2022dynamics,mukherjee2024influence}. We observe a common trend where the width of the distribution spans several orders of magnitude. This introduces a hierarchy of time scales in the cluster description that explains the emergence of glassy dynamics. 
However, a direct correlation between the distribution width and slow relaxation has not been fully explored. To address this, we derive a functional form of the standard deviation $\sigma_J$ for the distribution $\log J_{\rm NN}$, as a function of both dimension $d$ and power-law exponent $\alpha$ in the regime of very low filling fraction (see SM \cite{sup:info} for details)
\begin{equation}
	\frac{\sigma_{J}}{J_0} = \frac{\pi}{\sqrt{6}}\frac{\alpha}{d} .
    \label{sigma_alpha_d}
\end{equation}
This result is consistent with the numerically obtained $\sigma_J$ at $f\lesssim 10^{-3}$, shown in \fref{Fig1}~(c) for $\alpha=3$. As $f$ increases, $\sigma_J$ in \fref{Fig1}~(c) gradually decreases due to the reduced number of possible nearest-neighbor configurations. 
The expression \bref{sigma_alpha_d} serves as a crucial link between previous studies of glassy dynamics and our proposal for a unifying framework here. 

To systematically establish a connection between 
$\sigma_J$ and glassy relaxation,  we investigate the dynamics at very low filling fractions in three regimes of the XXZ model: (a) XY ($\Delta=0$), (b) Ising ($\Delta \gg 1$) and XXZ ($\Delta=0.6$), for various exponents $\alpha=2,3,...,6$. Given that \eref{sigma_alpha_d} implies $\sigma_J \propto \alpha/d$, and the well-established relation $\beta_{\rm pair} = d/\alpha$ for the Ising model \cite{schultzen2022glassy}, we thus investigate the dependence of $\beta$ on $1/\sigma_J$ for each case in \fref{Fig2}~(a-c). 
In all cases, we observe the stretch exponent is inversely proportional to the standard deviation of distribution i.e.~$\beta \propto 1/\sigma_J$ in the regime of $J_0/\sigma_J<1$, with minor variations in slopes depending on $d$ and $\Delta$. This leads us to propose an empirical expression for the stretch exponent of the form 
\begin{equation}
    \beta=g(\Delta,d)J_0/\sigma_J. 
    \label{beta_g_sigma}
\end{equation}
This leads to two key arguments that both align with and extend beyond the pair model predictions. First, the standard deviation of the disorder distribution $(\sigma_J)$ fundamentally governs $\beta$. 
Secondly, we assert that $\sigma_J/J_0 \gtrapprox 1$ provides a minimum condition required to observe glassy dynamics. 

We determine the behavior of $g(\Delta,d)$ by analyzing the dependence of $\beta$ with the model-dependent parameter $\Delta$, as shown in the inset of \fref{Fig2}~(d) for various $\alpha$. It is evident that each $\alpha$ results in a nearly distinct value of $\beta$. However, rescaling $\beta$ by $\sigma_J$, such that $\beta \sigma_J/J_0=g(\Delta,d)$ according to \bref{beta_g_sigma}, approximately collapses the data onto a single curve, as shown in \fref{Fig2}~(d-f) for 1D, 2D and 3D, respectively. Thus, the influence of $\Delta$ and the array dimension $d$ remains weak, causing only a $15$-$20\%$ variation in $g(\Delta,d)$. We attribute the majority of this scatter to the effects of finite filling fraction (see SM for additional discussion). The trend of $g(\Delta,d)$ with $\Delta$ aligns with previous observations for $\alpha=6$ and $d=3$ in Ref.~\cite{schultzen2022semiclassical}, as well as $\alpha=3$ and $d=1,2$ at $\Delta=0$ investigated in Ref.~\cite{mukherjee2024influence}.

\begin{figure}[tb]
	\centering
	\epsfig{file=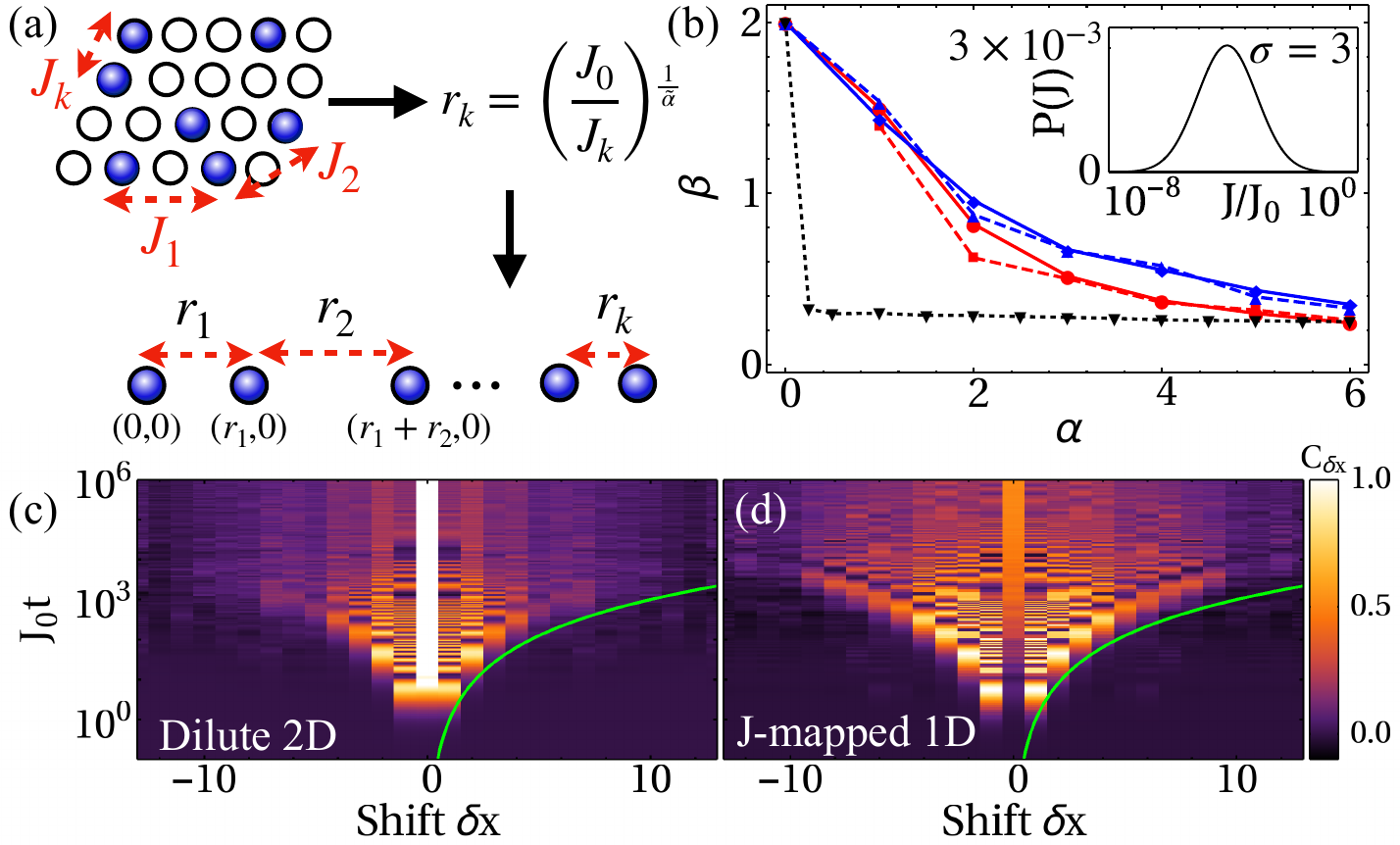,width= \linewidth}
	\caption{ (a) Illustration of $J$-mapping method (see text), where the nearest neighbor distribution of disordered 2D arrays is mapped into a deterministically positioned 1D array. (b) Variation of stretch exponent $\beta$ with power-law exponent $\alpha$ in 2D dilute array (red solid circle) and 3D dilute array (blue solid diamond), compared to J-mapped 1D array (dashed lines of same color) with fixed exponent $\tilde\alpha=3$. The black inverse-triangle dashed line indicates a 1D array for a log-normal distribution with $\sigma_J=3$ (see inset). (c-d) Correlation $C(\delta x,t)$ as a function of rescaled time $J_0t$ and intersite distance $\delta x$ in (c) dilute 2D array with $f=0.01$ and $L=28$ and (d) $J$-mapped 1D array.
		\label{Fig3}}
\end{figure}

To demonstrate the generality of our framework, i.e., how \eref{beta_g_sigma} applies beyond the touchstone problem of dilute arrays, we propose a \textit{$J$-mapping} (distribution-mapping) method, which can be used to map an arbitrary distribution of interactions $P(J)$ to a 1D array with power-law exponent $\tilde\alpha$. Our method is based on the systematic construction of a 1D array with nearest-neighbor spacings chosen to replicate a set of interactions sampled from $P(J)$. Concretely, we define the $k$-th spin separation as $r_k = (J_0 / J_k)^{1/\tilde\alpha}$, with $J_k$ randomly sampled according to $P(J)$ \cite{yoshida2024proposal}. In \fref{Fig3}~(a), we illustrate the J-mapping for a 2D array, placing spins sequentially at $(0,0)$, $(r_1,0)$, $(r_1 + r_2, 0)$, and so on.


We apply the $J$-mapping method to a log-normal distribution, where the logarithm of the random variable $J$ follows a normal distribution, i.e.,
\begin{equation}
	P(J) = \frac{1}{J\sqrt{2\pi\sigma_{J}^2}} \exp\left[ -\frac{\left( \log(J) - \mu_J \right)^2}{2\sigma_J^2} \right],
\end{equation}
where $\mu_J$ and $\sigma_J$ are the mean and standard deviation, respectively [see inset of \fref{Fig3}~(b)].
We map $P(J)$ to a 1D array, with fixed $\tilde\alpha=3$ and varying $\sigma_J$, and simulate the quench dynamics. Figure~\ref{Fig2}~(a-c) shows the variation of the extracted $\beta$ with $1/\sigma_J$. We observe the same functional scaling of $\beta$ as for the previous dilute arrays [i.e., matching Eq.~(4)].

Our analysis of log-normal disorder reveals that the dependence of $\beta$ on $\sigma_J$ provides a powerful tool for tuning glassy dynamics in quantum systems with controllable disorder. Using this, we apply our generalized framework to map $P(J)$ across systems with different dimensionalities and engineer glassy dynamics in quasi-ordered arrays.


    	%
%
	%

\ssection{Tunable relaxation in 1D arrays}
We apply J-mapping to emulate the dynamics of (i) 2D and (ii) 3D dilute arrays in bespoke 1D chains. This procedure enables us to demonstrate that the dynamics of numerous disordered systems, across various dimensionality and interaction range, can be effectively emulated using a carefully engineered 1D array.

We start with a dilute 2D array of $N=8$ qubits with filling fraction $f=0.01$.
In \fref{Fig3}(b), we compare the stretch exponents obtained from dynamics in the 2D dilute array (red solid circle) for various $\alpha$ and a $J$-mapped 1D array (red dashed square) with fixed $\tilde\alpha=3$. Overall, we observe good agreement between the stretch exponents from the J-mapped 1D array and the dilute 2D array. In the regime of small $\alpha \lesssim 2$ we observe minor discrepancies due to the influence of long-range interactions in the system, where spins beyond the nearest-neighbors begin to contribute to the dynamics. Thus our mapping, which only replicates the correct nearest-neighbour interactions, becomes inadequate.
We discuss this further in the SM, which also includes results using a mapped power-law exponent $\tilde\alpha = \alpha$ for the $J$-mapped 1D array. This approach substantially mitigates the discrepancy and effectively captures the long-range behavior observed in 2D systems with $\alpha \lesssim 2$.

    
In the case of the 3D dilute array (blue solid diamond), we observe a similar agreement with the J-mapped 1D array (blue dashed triangles) in the regime of $\alpha>2$, while we again observe minor discrepancies at small $\alpha$. Interestingly, in the context of the log-normal distribution with fixed $\sigma_J=3$ and varying $\tilde\alpha$, we demonstrate that the resultant dynamics in the system exhibit a near constant stretch exponent independent of $\tilde\alpha$, as shown in \fref{Fig3}~(b) with black-dashed inverse-triangle line. This suggests that specific scenarios can be designed where the influence of dimensionality and interaction range on the dynamics can be made negligible.

\begin{figure}[tb]
	\centering
	\epsfig{file=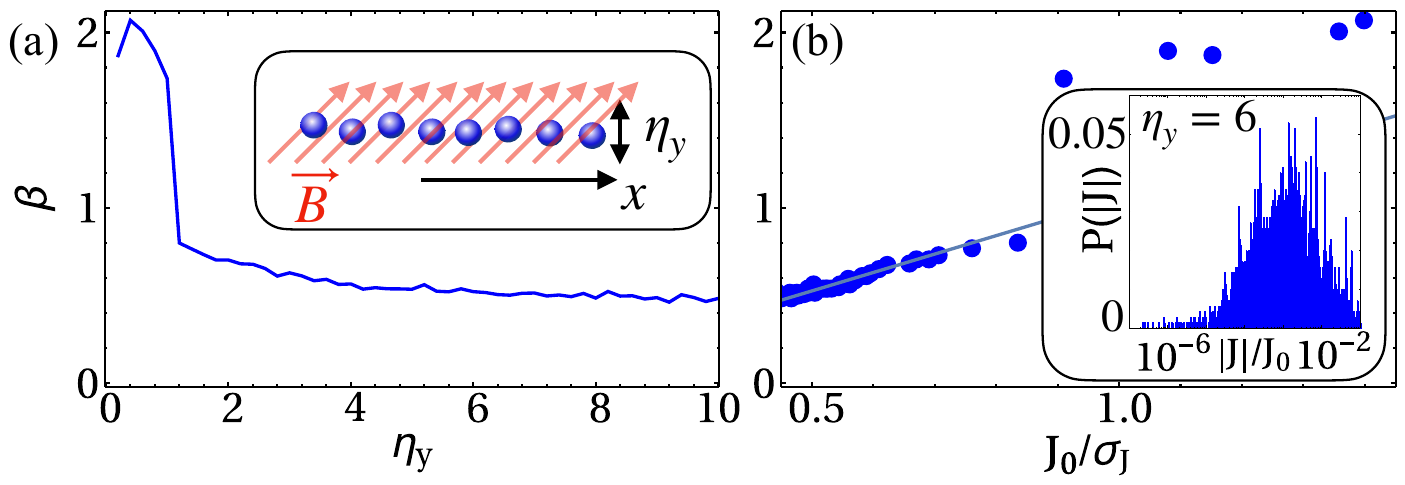,width= \linewidth}
	\caption{(a) Variation of the stretch exponent $\beta$, in a quasi-ordered 1D array, with the strength of the Gaussian fluctuations $\eta_y$ in the transverse direction (as inset for illustration). Red arrows indicate the direction of the magnetic field $\vec{B}$ to manipulate the quantization axis and introduce anisotropic dipolar interactions. (b) Variation in the stretch exponent with the inverse standard deviation $J_0/\sigma_J$ of the distribution of pair interactions (see example for $\eta_y = 6$ in inset). The blue line indicates predicted $g(\Delta,d)/\sigma_J$ for $\Delta = 0$ and $d = 1$. 
		\label{Fig4}}
\end{figure}
Finally, to assess the understanding provided by our framework beyond one-body observables, we investigate the dynamics of the site-resolved correlation function, $C(\mathbf{\delta r},t) = \mathcal{N}\sum_\mathbf{r} (\langle \hat{n}^{\leftarrow}_{\mathbf{r}} \hat{n}^{\leftarrow}_{\mathbf{r}+\mathbf{\delta r}}\rangle - \langle \hat{n}^{\leftarrow}_{\mathbf{r}} \rangle \langle \hat{n}^{\leftarrow}_{\mathbf{r}+\mathbf{\delta r}}\rangle)$ \cite{christakis2023probing,mukherjee2024influence}. Here, $\langle \cdot \rangle$ denotes averaging with respect to the quantum state and disorder, and $\hat{n}^{\leftarrow}_{\mathbf{r}}=1/2(\hat{\mathbb{I}} - \hat{\sigma}_x)_\mathbf{r}$ at position $\mathbf{r}$. We compare the propagation of correlations in a dilute 2D array, as shown in \fref{Fig3}~(c), to a $J$-mapped 1D array in (d), both with $\alpha=\tilde\alpha=3$. The green line in each panel indicates identical sub-ballistic spread of correlations at short times within a region $\propto (Jt)^{1/3}$. At long times, when disorder dominates the correlation dynamics \cite{mukherjee2024influence}, we observe that the $J$-mapped array similarly features an expected arrest in the spread of correlations, though the correlation width is about a factor of $2.5$ broader. Nevertheless, in the SM we show it retains the same $1/\sqrt{f}$ scaling observed in the dilute 2D array \cite{mukherjee2024influence}, demonstrating the robustness of the mapping in capturing the overall dynamics. We attribute the prefactor discrepancy to next-nearest-neighbor interactions \cite{mukherjee2024influence}.

With the insight provided by our unifying framework, we are also able to develop scenarios where glassy relaxation can emerge in quasi-ordered arrays.
This setup can be realized in experiments with a regularly spaced array along the axial direction, but random positioning along the transverse direction, as shown in the inset of \fref{Fig4}~(a). To exacerbate the interaction disorder, we consider anisotropic dipolar exchange interactions described by $J_{ij}=J_0(1-3\cos^2\theta_{ij})/|\mathbf{r}_{ij}|^3$, where $\theta_{ij}$ can be tuned through the orientation of an applied magnetic field relative to the axial direction. In this example, we assume the field is oriented at the magic angle of the array, i.e.~at $55$ degrees from the axial direction. This allows us to observe glassy features with relatively small transverse fluctuations.
    
Our design results in a distribution of interactions between strongly interacting pairs (not necessarily physical nearest neighbors) spanning several orders of magnitude, as shown in the inset of \fref{Fig4}~(b). 
The statistical properties of the distribution ($\mu_J$ and $\sigma_J$) can be controlled by adjusting the amplitude of the transverse position fluctuations $\eta_y$ of the tweezers. In \fref{Fig4}~(a), we show the variation in $\beta$ with $\eta_y$, showing that $\beta$ can be tuned over a broad range, from 2 down to 0.4. We note that for $\eta_y \gg 1$ the extracted exponent eventually saturates near $0.4$, consistent with the stretch exponent expected for a 2D dilute array with XY interactions. Furthermore, the variation of $\beta$ follows the expected $g(\Delta,d)J_0/\sigma_J$ scaling in the regime $\sigma_J/J_0>1$ for $\Delta=0$ and $d=1$ [see \fref{Fig4}~(b)]. 

\ssection{Outlook}
In future work, our framework could be extended to incorporate correlated positional disorder, where the spins are distributed using a non-Poissonian distribution. Such correlated disorder could break the validity of descriptions based on very small clusters and potentially give rise to more complex many-body relaxation. Moreover, recent advances in dynamical reconfiguration capabilities could be leveraged to investigate time-dependent positional disorder, enabling dynamical control of the propagation of correlations and entanglement. 

    \begin{acknowledgments}
        \textit{Acknowledgments:} We acknowledge stimulating discussions with Martin G{\"a}rttner. This material is based upon work supported by the Air Force Office of Scientific Research under Grant No. FA9550-22-1-0335. The computing for this project was performed at the OU Supercomputing Center for Education \& Research (OSCER) at the University of Oklahoma (OU).
    \end{acknowledgments}
    
	
	%

	\newpage
	\section{Supplemental Material}

    \section{Mean and variance of distribution of nearest-neighbor interactions \label{app:mean_variance}}
The analysis of glassy dynamics in the main text is based on the identification of key traits in the distribution of nearest-neighbor interactions in positionally disordered spin models. Here, we derive the analytic expressions for the mean and variance of the nearest neighbor distribution. 

The distribution of nearest neighbors with spin separated by a distance $r$ (in units of $a_0$) in $d$-dimensional dilute arrays is expressed in the form \cite{mukherjee2024influence} 
\begin{eqnarray}
    P_{\mathrm{1D}}(r) &=& 2f(1-f)^{2r-1} , \\
    P_{\mathrm{2D}}(r) &=& 2\pi r f(1-f)^{\pi(r-1)^2} , \\
    P_{\mathrm{3D}}(r) &=& 4\pi r^2 f (1-f)^{4\pi(r-1)^2/3} , 
\end{eqnarray}
where $f = N/L^d$ is the filling fraction associated with $N$ atoms distributed over $L^d$ sites.  
We use these expressions to derive statistical features of the associated distribution of interactions, such as $\sigma_J$ and $\mu_J$ in terms of dimensionality $d$ and power-law exponent $\alpha$. To do this, it is useful to relax the underlying framework of a unit-spaced array (i.e.~regular lattice of sites with associated lattice constant $a_0$) and instead work with arbitrarily positioned tweezers. Specifically, if we assume $f$ is the average number of atoms per unit length/area/volume, then the nearest-neighbor spacing distributions are given by,
\begin{eqnarray}
    \tilde{P}_{\mathrm{1D}}(r) &=& 2fe^{-2fr} , \\
    \tilde{P}_{\mathrm{2D}}(r) &=& 2\pi r e^{-\pi r^2} , \\
    \tilde{P}_{\mathrm{3D}}(r) &=& 4\pi r^2 e^{-4\pi r^2/3} .
\end{eqnarray}
From these we can compute the moments of the associated interaction strength $J = J_0/r^{\alpha}$:
\begin{eqnarray}
    \frac{\mu_{J,\mathrm{1D}}}{J_0} = \alpha\gamma + \alpha\log(2f) \quad \text{and} \quad \frac{\sigma_{J,{\mathrm{1D}}}}{J_0} = \frac{\alpha \pi}{\sqrt{6}}, \\
    \frac{\mu_{J,\mathrm{2D}}}{J_0} = \frac{\alpha\gamma}{2} + \frac{\alpha}{2} \log(\pi f) \quad \text{and} \quad \frac{\sigma_{J,{\mathrm{2D}}}}{J_0} = \frac{\alpha \pi}{\sqrt{24}}, \\
    \frac{\mu_{J,\mathrm{3D}}}{J_0} = \frac{\alpha\gamma}{3}+  \frac{\alpha}{3}\log(4\pi f/3) \quad \text{and} \quad \frac{\sigma_{J,{\mathrm{3D}}}}{J_0} = \frac{\alpha \pi}{\sqrt{54}} ,
\end{eqnarray}
where $\gamma \approx 0.577...$ is Euler's number. From these expressions, we identify a key result: The standard deviation scales as,
\begin{equation}
    \frac{\sigma_{J}}{J_0} = \frac{\pi}{\sqrt{6}}\frac{\alpha}{d} ,
\end{equation}
in $d$ dimensions, which we find to be consistent with numerically obtained $\sigma_J$ at $f\lesssim 10^{-3}$ in \fref{Fig1}~(c) of the main text. 

    \section{Dependence of $g(\Delta,d)$ on filling fraction}

\begin{figure}[htb]
	\centering
	\epsfig{file=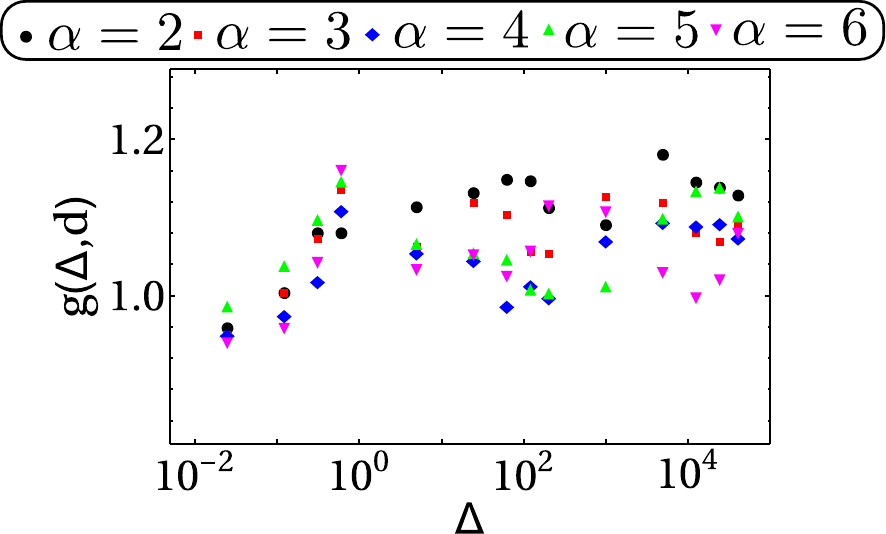,width= 0.65\linewidth}
	\caption{ Variation in $g(\Delta,d)=\beta \sigma_J/J_0$ with anisotropy parameter ($\Delta$) in 1D disordered array with $f=0.1$ for different power-law exponents $\alpha$.
		\label{Fig_appendix_g_Delta}}
\end{figure}
    In the main text, we examine the behavior of $g(\Delta,d)$  as a function of the anisotropy parameter $\Delta$ and dimension $d$ for a filling fraction $f=0.01$. For reference, we present the values of $g(\Delta,d)$ for $d=1,2,3$ and $\Delta$ in \Tref{tab:my_label}, obtained from slopes in Fig.~2 (a-c). In Fig.~2 (d-f), we observed that the curves do not collapse perfectly onto each other. We attribute this scatter to the finite filling fraction. To illustrate this, in \fref{Fig_appendix_g_Delta} we show results for $g(\Delta,d)$ obtained with a larger $f = 0.1$. The color scheme used is consistent with that in Fig.~2(d–f). Here, we observe that the scatter is relatively worse in the regime $\Delta>1$ across different power-law exponents $\alpha$, which supports our reasoning. 

\begin{table}[htb]
 \caption{Values of $g(\Delta,d)$ for various $\Delta$ and $d$ used in Fig.~2 (a-c)}
    \begin{center}
\begin{tabular}{|c | c| c |c|} 
 \hline
  & XY ($\Delta=0$) & XXZ ($\Delta=0.6$) & Ising ($\Delta\gg 1)$ \\ [0.5ex] 
 \hline
 1D & 1.05152 & 1.2458 & 1.29921 \\ 
 \hline
 2D & 0.942196 & 1.06197 & 1.15824 \\
 \hline
 3D & 0.969538 & 1.12121 & 1.20124 \\ [1ex] 
 \hline
\end{tabular}
\end{center}
    \label{tab:my_label}
\end{table}

    \section{J-mapped arrays}

\begin{figure}[htb]
	\centering
	\epsfig{file=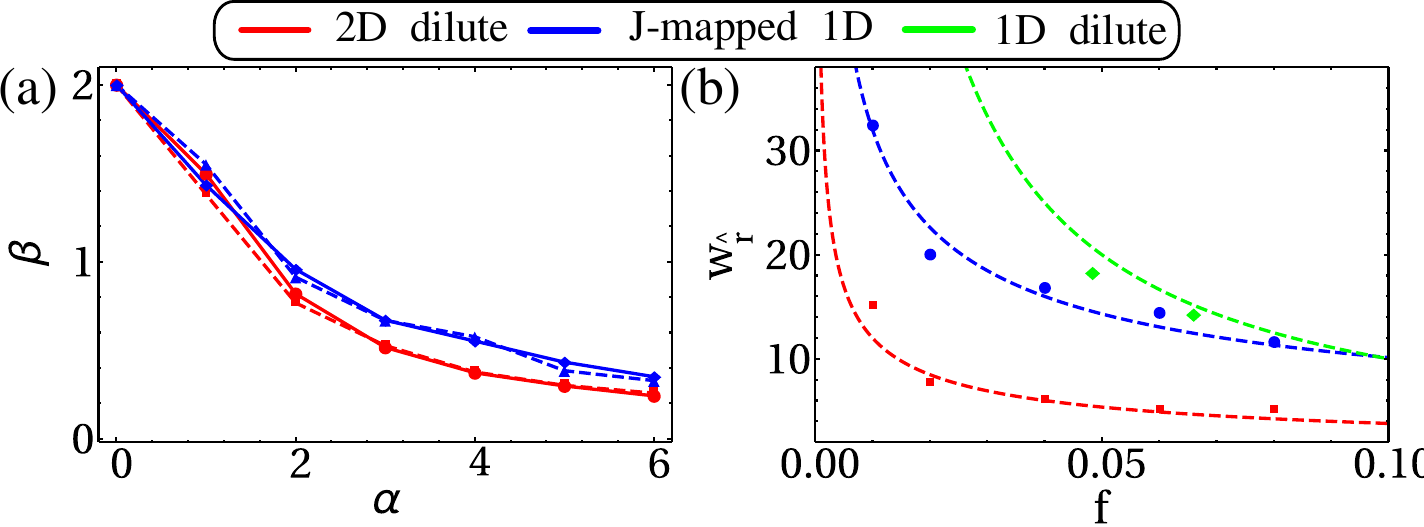,width= \linewidth}
	\caption{  (a) Variation of stretch exponent $\beta$ with power-law exponent $\alpha$ in 2D dilute array (red solid circle), compared to the J-mapped 1D array with fixed $\alpha=3$ (blue squares). (b) Long-time correlation width $w_{\hat{r}}$ as a function of filling fraction $f$ in a dilute 2D array (red-square), J-mapped 1D array (blue circles) and dilute 1D array (green diamonds) with effective filling fraction of the J-mapped array. Dashed lines indicate $1/\sqrt{f}$ (red and blue) and $1/f$ (green) scaling.
		\label{Fig_appendix_2}}
\end{figure}

We demonstrate J-mapping for both 2D and 3D dilute arrays in the main text. Here, we present an additional example with power-law exponent of $1$D bespoke array chosen to match that of the emulated system, i.e., $\tilde\alpha = \alpha$. We compare the resulting stretch exponents to those from 2D dilute arrays with $\alpha = 0, 1, \dots, 6$ in \fref{Fig_appendix_2}~(a). Compared to Fig.~\ref{Fig3}~(b) of the main text, we observe good agreement across all values of $\alpha$ and, in particular, only marginal discrepancies at small $\alpha \lesssim 2$. 



		%
%


    To compare the correlation dynamics of a 2D dilute array with the corresponding J-mapped 1D array, we consider two key aspects: (i) the initial growth of correlations and (ii) the correlation width at long times. The initial growth is primarily independent of disorder \cite{mukherjee2024influence} and dimensionality, and is therefore accurately captured by the J-mapped array, as shown in Fig.~3(c–d). In contrast, the long-time correlation width includes more complex many-body effects. In \fref{Fig_appendix_2}~(b), we compare the correlation widths at long times and find that, although they differ by an overall factor of $2.6$, both exhibit the same $1/\sqrt{f}$ scaling, a signature behaviour of 2D dilute arrays. This consistency suggests that the J-mapping method remains largely effective even when capturing two-body observables. Furthermore, for reference, we show the correlation width for 1D dilute array with effective filling fraction from $f_{\rm eff}$ obtained from the J-mapped array, and it can be seen that in this case the correlation width follows a distinct $1/f_{\rm eff}$ scaling \cite{mukherjee2024influence}.

\begin{figure}[htb]
	\centering
	\epsfig{file=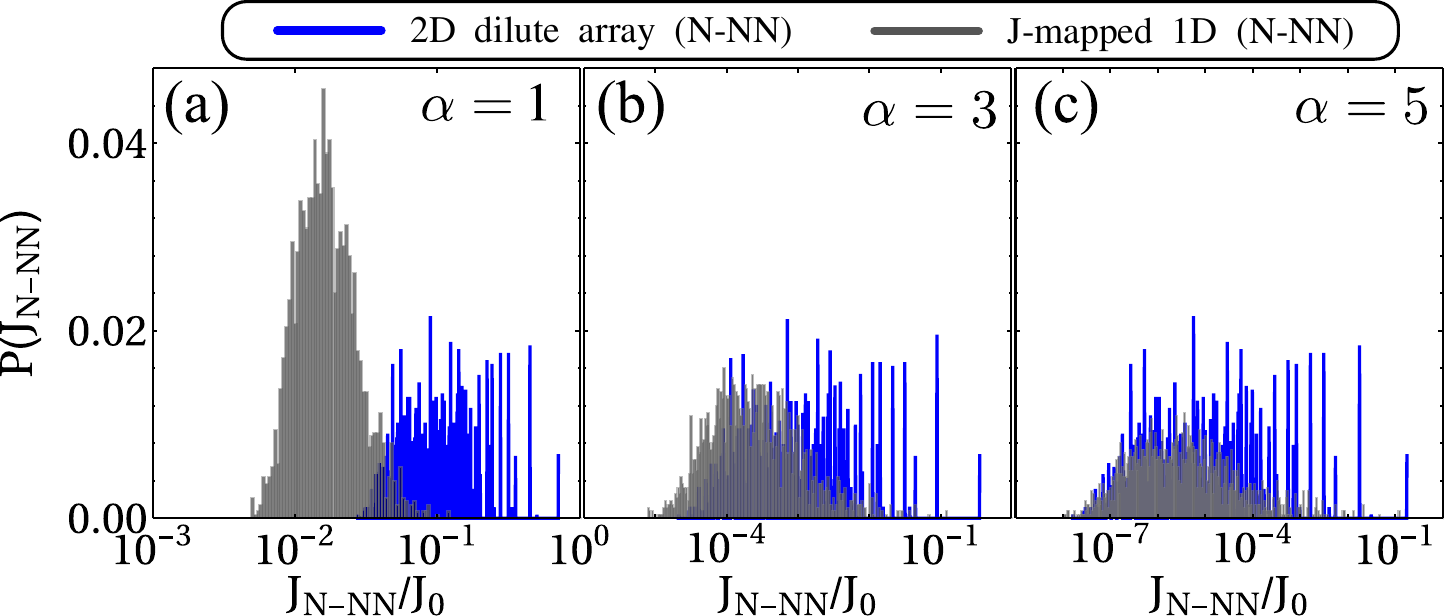,width= \linewidth}
	\caption{ Next-nearest neighbor (N-NN) distribution for power law exponent (a) $\alpha=1$, (b) $\alpha=3$, and (c) $\alpha=5$ in 2D dilute array (blue) and its corresponding J-mapped 1D array with $\tilde\alpha=3$ (gray) for parameters in Fig.~3.
		\label{Fig_appendix}}
\end{figure}
    \section{Next-nearest neighbor (NNN) distribution}\label{app:NNN}

    In the main text, we discussed that the J-mapped 1D array with a fixed power-law exponent $\tilde\alpha = 3$ is unable to fully capture the dynamics in regimes $\alpha <\tilde\alpha$. We attributed this to the fact that long-range interactions, and thus next-nearest-neighbour (N-NN) interactions, should begin to play a non-negligible role in the dynamics of the $2$D and $3$D dilute arrays. Specifically, our $J$-mapped $1$D arrays are constructed by the sequential placement of spins according to the distribution of interactions between strongly coupled [i.e., nearest-neighbour (NN)] pairs of the original $2$D or $3$D array. As the location of each new spin only depends on the position of the directly preceding partner, and not on the location of every other spin in the array, there is no expectation for N-NN interactions to match.

    In \fref{Fig_appendix}(a–c), we present the N-NN interaction distributions for a 2D dilute array (blue) with power-law exponents $\alpha = 1, 3, 5$, alongside those obtained from the corresponding J-mapped 1D arrays. We observe two key characteristics. First, as $\alpha$ increases, the overall magnitude of the N-NN contributions decreases, indicating a reduced influence on the relaxation dynamics. Second, for larger values of $\alpha$, such as $\alpha = 5$, the N-NN distributions of the 2D and $J$-mapped 1D systems show significant overlap, suggesting that even if N-NN interactions become relevant to the dynamics they will be reasonably captured by the $J$-mapped array. In contrast, as $\alpha$ decreases, the discrepancy between the N-NN distributions becomes more pronounced at the same time as they become more important for the relaxation dynamics. This disagreement, in the regime of small $\alpha$, where the N-NN contributions are most significant, is responsible for the deviations observed in Fig.~3 (b) at $\alpha<3(=\tilde\alpha)$.

    We note that this analysis may also be relevant to the discrepancies in observed correlation widths at long times, as discussed in Fig.~\ref{Fig3} of the main text and Fig.~\ref{Fig_appendix_2} discussed in this SM. These datasets correspond to $\tilde\alpha = \alpha = 3$, for which the N-NN interaction distributions for a $2$D dilute array and $J$-mapped counterpart are shown in Fig.~\ref{Fig_appendix}(b). There, we observe minor disagreements between the distributions. Given that the interactions between clusters of three spins are crucial for the behavior of the correlations at long times, as detailed by us in Ref.~\cite{mukherjee2024influence}, we expect this disagreement to explain the small quantitative discrepancies in the widths of the correlations of the $2$D dilute and $J$-mapped arrays. 

\end{document}